\documentclass[12pt]{iopart}
\usepackage{graphicx}
\begin{document}
%====================================================================%

\title{An {\it ab initio} study of a field-induced position change of a C$_{60}$ molecule adsorbed on a gold tip} 

\author{R.~Stadler$^1$ \footnote[3]{To whom correspondence should be addressed (stadler@fysik.dtu.dk)}, S.~Kubatkin$^2$, and T.~Bj{\o}rnholm$^3$}
\address{$^1$Center for Atomic-Scale Materials Physics, Department of Physics, NanoDTU, Building 307, Technical University of Denmark, DK-2800 Lyngby, Denmark}
\address{$^2$Department of Microtechnology and Nanoscience (MC2), Fysikgrand 3, Chalmers University of Technology, S-41296, G\"{o}teborg, Sweden}
\address{$^3$Nano-Science Center and Department of Chemistry, University of Copenhagen, Universitetsparken 5, DK-2100, Copenhagen, Denmark}

\begin{abstract}
Recent I/V curve measurements suggest that C$_{60}$ molecules deposited in gold nanojunctions change their adsorption configuration when a finite voltage in a 2-terminal setting is applied. This is of interest for molecular electronics because a robust molecular transistor could be based on such junctions if the mechanism of the process is understood. We present density functional theory based plane wave calculations, where we studied the energetics of the molecule's adsorption under the influence of an external field. Particular emphasis was placed on investigating a possible lightning rod effect which might explain the switching between configurations found in the experiments. We also analyze our results for the adsorption energetics in terms of an electrostatic expression for the total energy, where the dependence of the polarizability of the junction on the position of the C$_{60}$ molecule was identified as a crucial property for the field induced change of adsorption site.
\end{abstract}

\section{Introduction}\label{sec:intro}

A tremendous effort has been put into studying the adsorption of fullerene molecules on noble metal surfaces due to the many conceivable technological applications of thin films of such adsorbates as lubricants or in nonlinear optical devices or superconductors \cite{review}. Renewed interest has been triggered by two proposals for the C$_{60}$/Au system to be used as single transistors in the emerging field of molecular electronics \cite{molelec}. The mechanism of the first transistor concept was based on nanomechanical oscillations of the fullerene molecule in the trapping potential of the metallic lead \cite{park}, whereas in the second case a scanning tunneling microscope (STM) was used as a gate, which operates by compressing the molecule and thereby changing its conductance \cite{joachim}. Theoretical studies exist for the adsorption energetics of C$_{60}$ monolayers on Au(111) surfaces \cite{wang}, zero bias charge transfer effects in this system \cite{louie} and direct simulations of STM scans \cite{jimenez}.\\
Recently, new measurements have been reported \cite{kubatkin} of single C$_{60}$ molecules trapped in nanojunctions prepared by quench condensation in UHV of gold source and drain electrodes on top of an aluminium gate covered with a thin oxide. These measurements reveal a strong interaction between the fullerene molecule and the gold electrodes resulting in the absence of the Coulomb blockade effects observed by other groups \cite{park}. Additionally the measurements indicate the presence of two metastable structures in the junction defined by the leads and the adsorbate, where a switching between them can be induced by applying a voltage of $\sim$ 50 mV. Danilov et al. \cite{kubatkin} suggest a variety of different mechanisms for this field-induced reversible switching process, where in all of them the properties of the surface/adsorbate interface play the key role, because the molecule itself does not have metastable conformations. One of these mechanisms is built on the assumption that the strength of the applied electric field is varying at the Au surface due to the significant corrugation of its granular structure. The molecule would then change its position to a point of highest field strength in its vicinity when a voltage is applied, which would not necessarily be the most stable adsorption site in the absence of the field. To be more specific, one would expect the molecule at zero bias to adsorb on the side of a protrusion formed by a gold grain, because there it would have more Au atoms available to bond to. At the critical voltage it would move to the tip of the protrusion, where the field strength would be highest. Since the two positions could correspond to different distances to the electrode on the other side of the $\sim$ 2 nm gap, the configuration switching could explain the variation in conductance found in the experiments \cite{kubatkin}.\\
Such a mechanism is plausible because the local enhancement of an electric field has been found to play an important role in optical spectroscopy (e.g. local fluorescence of Raman spectra) \cite{optical} and the field dependent adsorption site preferences of adatoms influences the properties of field ion emitters \cite{tsong}. In recent density functional theory (DFT) studies the effect of the local field of a STM tip on the energetics and stability of competing surface structures has been investigated, where effects such as 'dimer switching' \cite{sidimer} and changes in adsorbate desorption barriers or STM detectabilities \cite{siadsorb} were found for silicon surfaces.\\
Our aim in the current work is to investigate the adsorption preferences of a C$_{60}$ molecule on a Au (111) surface with a protrusion of 4-5 atoms in dependence on an external electric field by using a DFT plane wave approach \cite{dacapo}. The paper is organized as follows: In the following section we introduce the computational details of our calculations and the atomic configurations of the investigated structures. In section 3 we report our main results for the field dependent total energies of the different structures and discuss their implications with respect to the experimental findings in Ref. \cite{kubatkin}. In section 4 we present an electrostatic analysis of our main results, where the important role of the polarizability of the interface between the surface and the molecule is highlighted. We conclude with a summary.

\section{Surface/adsorbate geometries and computational details}

Fig.\ \ref{fig1} shows the variety of the surface structures (A,B) and C$_{60}$ adsorption positions (1,2) investigated in this article. Additionally to the molecule and one Au tip on each side of our slab geometry, the corresponding supercells contain four Au layers in a (111) stacking with 4x4 periodicity within the surface plane and a $3 \times 3$ {\bf k}-point grid in the transverse Brillouin zone. For both surface structures the lateral cell size has been chosen such that the distance between the upmost Au tip atoms across the vacuum was 20 \AA . Whereas for the Au surface truncated bulk coordinates have been used, the positions of the carbon nuclei have been optimized for the isolated molecule within our DFT-plane wave approach \cite{dacapo} using ultrasoft pseudopotentials \cite{vanderbilt}, the PW91 functional for exchange and correlation \cite{pw91} and an energy cutoff of 340 eV for the plane wave expansion.\\
For both positions the surface/molecule distance has been varied perpendicular to the surface and for position 2 additionally parallel to the surface in order to ensure that all geometries correspond to local total energy minima at zero bias. In all geometries the closest C-Au distance was found to be in the range 2.75-3.1 \AA\ , which agrees well with the bonding range explored in Ref. \cite{jimenez}. It is apparent from previous DFT calculations on the C$_{60}$/Au(111) system \cite{wang,jimenez} that the binding energy between the molecule and the surface will also depend on the rotational orientation of the fullerene with respect to the surface atoms. However, due to the (compared with Refs. \cite{wang,jimenez}) increased structural complexity of our system including also a Au tip and because we are not interested in absolute values for the binding energies but rather in their changes induced by an electric field, we did not optimize our geometries with respect to this parameter. In both of our surface structures a pentagon of the C$_{60}$ molecule faces the Au tip atom in position 1 and the molecule is just laterally shifted for position 2 without any further rotation.\\
Our method for applying an electric field is to add a linear ramp to the external potential in the Kohn Sham equations for a self-consistent determination of the electronic degrees of freedom \cite{jan}, where the discontinuity of the potential for periodically repeated cells is placed in the middle of the vacuum at the plane with the smallest summed up electron density. The same technique has been previously employed for removing artificial dipole-dipole interactions occuring in a repeated slab scheme, where the unit cell has a dipole moment in the direction perpendicular to the surface \cite{method}. In this scheme there are two parameters defining the electric field, namely the potential difference at the discontinuity and the length of the unit cell in the direction perpendicular to the surface. The field strength is then just the quotient of the two.

\section{Field dependent total energies}

\begin{figure}
\includegraphics[width=1.0\linewidth,angle=0]{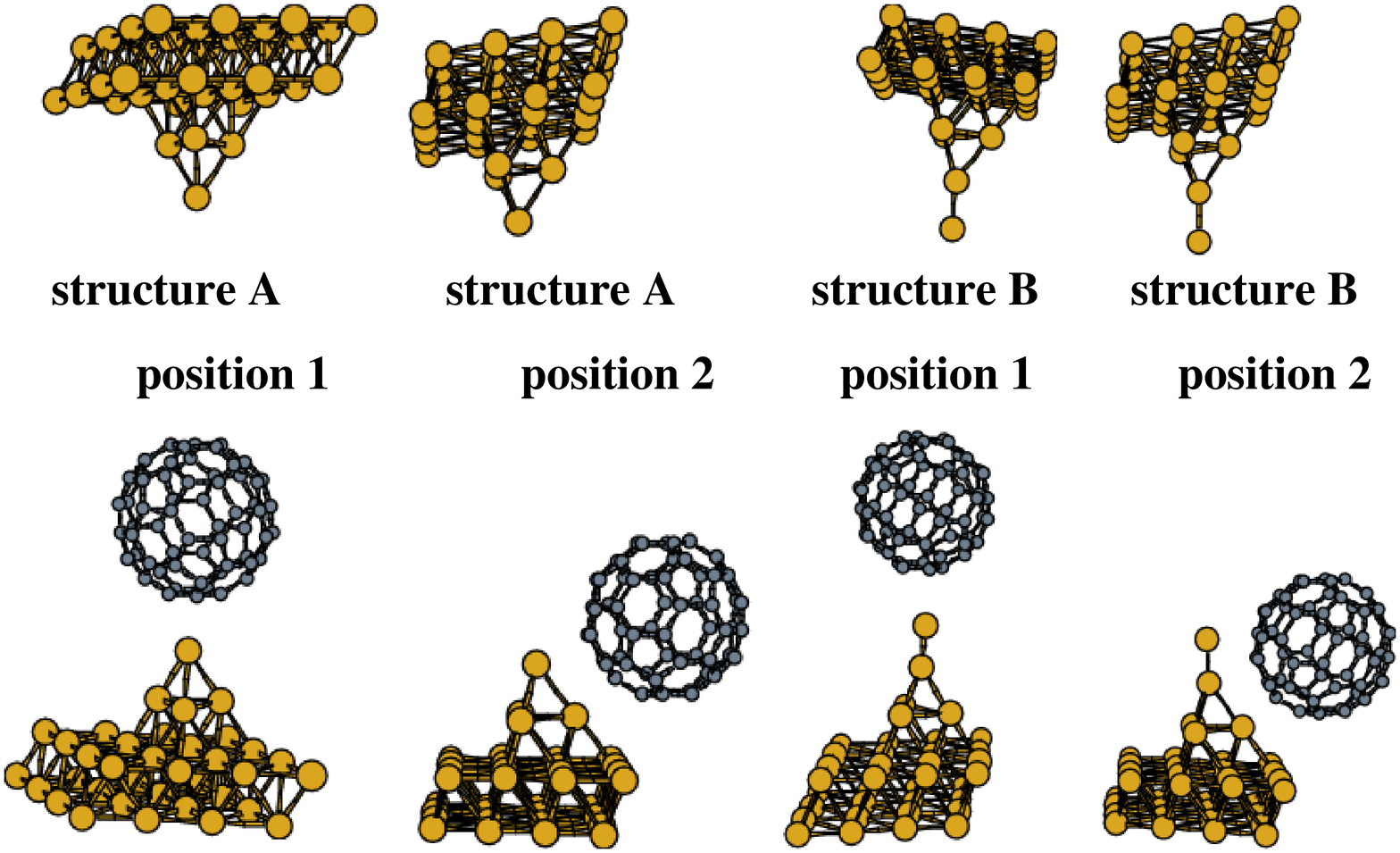}
\caption{\label{fig1}Structural overwiew of the four different surface/adsorbate geometries investigated in this paper. Structure A and B refer to different models for the Au tip with 4 and 5 atoms, respectively. Position 1 and 2 distinguish between the C$_{60}$ molecule being adsorbed on top or on the side of the tip, respectively.}
\end{figure}

\begin{figure}
\includegraphics[width=1.0\linewidth,angle=0]{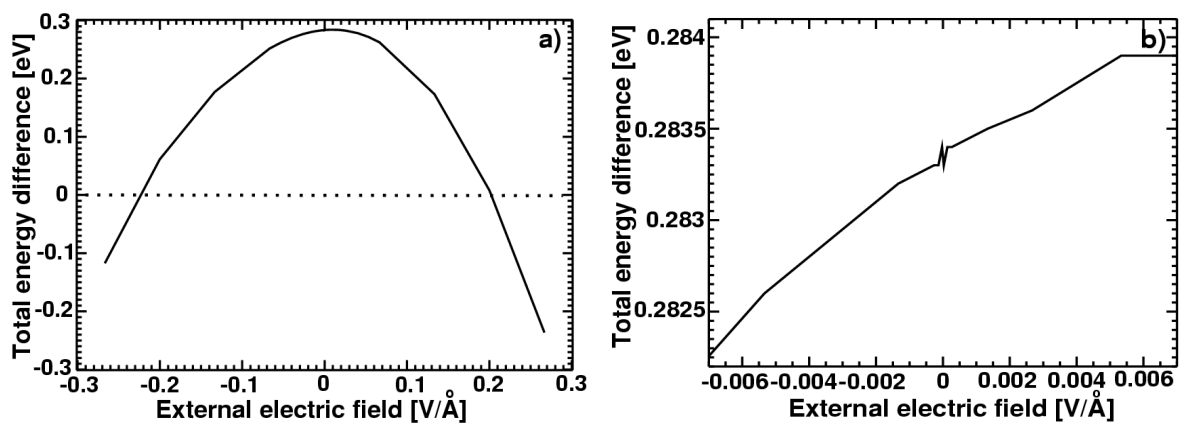}
\caption{\label{fig2}Total energy differences (position 1 - position 2) for structure A in dependence on an external elecric field for a) a rather large range of the field strength and b) within the linear regime around zero bias.}
\end{figure}

In Fig.\ \ref{fig2} we show the total energy difference between position 1 and 2 of structure A (see Fig.\ \ref{fig1}) in dependence on an external electric field. For our comparison it is crucial that an identical computational setup is used for the two geometries and only the molecular positions differ. Then possible sources of inaccuracy or ambiguity due to e.g. the limited and potentially varying number of Au layers in our cell can be assumed to cancel out, leaving only the energetic effect coming from the difference in the two surface/adsorbate interface structures. In Fig.\ \ref{fig2}a it can be seen that the fullerene molecule prefers to sit at the site of the tip (position 2), where its bonding to the surface is maximized due to the large number of nearby Au atoms. With an increase in the field, however, the situation reverses gradually and the molecule finally switches at an external field strength of $\pm$ 0.2 V/\AA\ to the on-top position of the tip (position 1), where the local field would be expected to be largest. \\
In the experiments of Ref. \cite{kubatkin}, however, the switching effect is observed at a voltage of 50 mV with the size of the gap between the electrodes being approximately 2 nm. This would result in a field strength of $\pm$ 0.0025 V/\AA\ , which is two orders of magnitude smaller than our calculated result. Additionally, the experimental data indicates (in combination with thermodynamic model assumptions \cite{per}) that the change in enthalpy induced by the switching voltage would be of the order of 5 meV and would have to occur in the linear regime in order for the model to reproduce the trends in the measured I/V curves. In Fig.\ \ref{fig2}b we zoom in on the linear regime of our calculated energy difference curve and find that even the assumption of a possible rigid shift of the curve, where it would cross the 'switching energy line' ($\Delta E_{tot}$=0) at lower field strength would not reconcile our theoretical results with the experimental findings. In our calculated function the entire linear regime covers only an energy range of 1.5 meV with $\sim$ 0.5 meV corresponding to $\pm$ 0.0025 V/\AA\ . \\
When directly comparing our calculations with the experiments, however, one has to keep in mind that there has to be a discrepancy in the spatial dimensions of the Au tip in relation to those of the fullerene molecule. In our calculations the C$_{60}$ molecule (having a diameter of $\sim$ 8 \AA\ ) interacts with a tip of less than 5 \AA\ length (for structure A). In the experiments the protrusion on the electrode would be made out of Au grains with diameters of up to a few nanometers \cite{grains}. This discrepancy could have an impact on the field induced adsorption energetics, since it is well known from optical spectroscopy and field emission that the local field enhancement factor at the tip of a protrusion can vary by up to a factor of 50-100 depending on its curvature and length \cite{curvature}.\\
Unfortunately we cannot increase the length of the tip drastically in our calculations without arriving at cell sizes which are computationally unfeasible for our DFT plane wave approach. So instead, for the sake of having two different tip lengths to compare, we enlarge the length of the tip by adding just a single Au atom (structure B in Fig.\ \ref{fig1}), which is equivalent to a length increase of about 50 percent. Such a surface structure is artificial in the sense that it would not be thermodynamically stable and the additional Au atom would migrate to the side of the tip instead of staying on top of it. The tip atom (which has only one Au neighbour) is also very reactive in structure B with regard to its bonding to the molecule, which means that the fullerene would always want to sit on top of the tip, even at zero bias. But since the main purpose of performing calculations for structure B is to study the effect of an elongation of the tip on the field dependence of our energetics (i.e. we are more interested in the curvatures of the functions in Figs.\ref{fig2}+\ref{fig3} than in absolute energy values), we shift its total energy difference curve rigidly for the comparison shown in Fig.\ \ref{fig3}, so that its maximum coincides with that of structure A. It can be seen that as expected the field dependence is much stronger for the increased tip height, which would reduce the switching threshold for the external field by $\sim$ 25 percent.

\section{Electrostatic analysis}

\begin{figure}
\includegraphics[width=0.8\linewidth,angle=0]{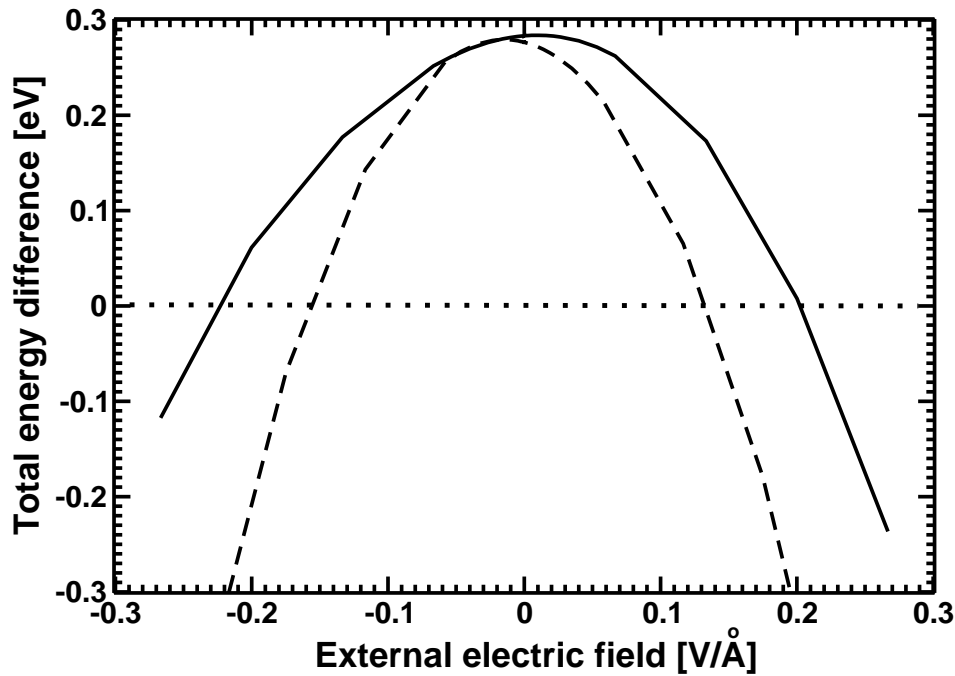}
\caption{\label{fig3}Total energy differences in dependence on an external electric field (see also Fig. 2) for structure A (solid line) and structure B (dashed line). The dashed curve has been shifted rigidly along the energy axis, so that the two maxima coincide (see text).}
\end{figure}

\begin{figure}
\includegraphics[width=1.0\linewidth,angle=0]{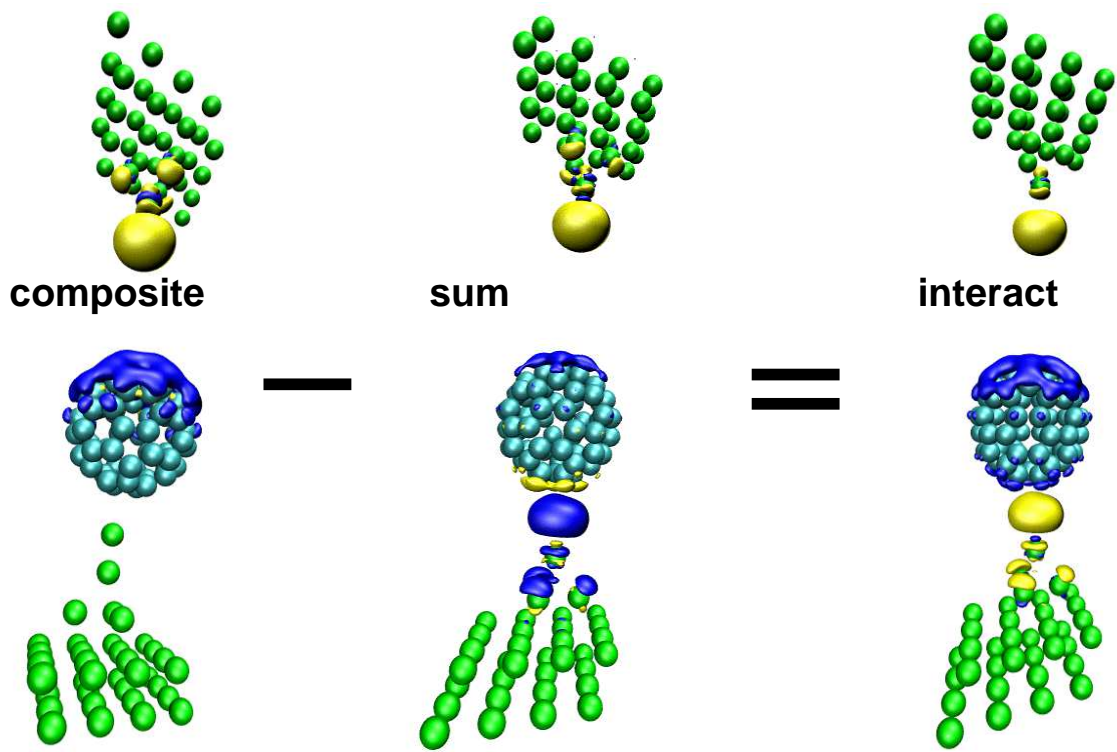}
\caption{\label{fig4}Differences in electron density for structure B position 1 (see Fig. 1) between an unbiased interface and one with an applied field with {\bf E}=-0.25 V/\AA\ . The areas of increased and decreased charge density due to the field are marked yellow and blue, respectively. The three pictures distinguish between the combined interface and the effect the field would have on the molecule and the Au slab separately (see text).}
\end{figure}

In order to analyze the structure dependence of the total energy curves in Figs. \ref{fig2}+\ref{fig3} further, it is useful to expand E$_{tot}$ as a second order polynomial of the strength of the electric field {\bf E},
\begin{equation}
E_{tot}=E_{tot,{\bf E}=0}-\mu_{perm}{\bf E}-\frac{1}{2}\alpha{\bf E}^2=E_{tot,{\bf E}=0}-(\mu_{perm}+\frac{1}{2}\mu_{ind}){\bf E}
\end{equation}
where the permanent dipole moment $\mu_{perm}$ is due to zero bias charge transfer, and the polarizability $\alpha$ (or induced dipole moment $\mu_{ind}$) characterize the response of the interface to the field in dependence on its geometry. In Table \ref{dipole.tab} we show numerical values for dipole moments and polarizabilities of all four geometries investigated in this article. There are two ways for calculating $\mu_{perm}$ and $\alpha$ from our DFT calculations, namely either directly from the charge densities at a given {\bf E} (which has been done for the main values in Table \ref{dipole.tab}) or from a polynomial fit of the energy curves in Figs. \ref{fig2} +\ref{fig3} (where numbers calculated that way are shown in the parantheses of Table \ref{dipole.tab}). The observation that both sets of numbers agree with each other in a rather satisfactory way suggests that the energetics we investigate is dominated by electrostatic effects and that exchange, correlation and kinetic energies play only a minor role.\\
It can also be seen that the numbers for the permanent dipole moments are rather small (at least compared with $\mu_{ind}$), and quite independent of the position of the fullerene molecule for a given surface (or tip) structure. It has to be noted, however, that for the linear regime (i.e. for very low biases) $\mu_{perm}$ will be in any case dominant and is therefore determining the slope in Fig. \ref{fig2}b. The polarizabilities and induced dipole moments (the latter calculated at {\bf E}=-0.25 V/\AA\ ) on the other hand show a very strong dependence on the position of the C$_{60}$ molecule. Whereas for both structures in position 2 (on the side of the tip) the induced dipole moment $\mu_{ind}^{comb}$ for the interface combining the molecule and the surface is only larger than $\mu_{ind}^{isol}$ for the isolated components by a factor of 1.2-1.4, it is more than twice its value in position 1 (on top of the tip). Additionally the values for $\mu_{ind}^{comb}$ reflect the trends in the slopes of the energy curves in Fig. \ref{fig3}, since its value is magnified with respect to $\mu_{ind}^{isol}$ by a factor of 1.44 for structure A and of 1.98 for structure B, respectively, when moving from position 1 to 2.\\
Finally, it is interesting to evaluate whether the differences between $\mu_{ind}^{comb}$ and $\mu_{ind}^{isol}$ have their origin in an enhancement of the local field at the position of the buckyball due to the presence of the tip (as it is commonly assumed in optical spectroscopy \cite{optical}) or whether they are due to field-induced charge transfer between the surface and the molecule (as it was suggested for adatoms on metal tips in the context of field ion emission \cite{tsong}). For this purpose we plot in Fig. \ref{fig4} the differences in electron density from our calculations at {\bf E}=-0.25 V/\AA\ and at zero bias, respectively, for position 1 of structure B, because for this geometry the difference of the induced dipole moments from the combined and isolated systems is the most pronounced. In the figure we compare the field-induced charge density change for the coupled interface (composite) with the sum of the changes in the fullerene molecule and the Au slab separately (sum) and show also the difference between the two (interact), which represents the field-induced electron transfer due to the interaction between the two subsystems. In this way we find that the molecule acts as an extension of the tip, where substantial movement of charge from the molecule ($\sim$ 0.5 electrons) to the slab occurs. With a reversal of the direction of the bias the direction of the charge movement would of course also be reversed. We conclude our analysis with the notion that our results cannot be explained in terms of just a local field enhancement due to the presence of the tip, and that charge transfer effects have to be explicitly accounted for in a realistic description of the systems under investigation.

\section{Conclusions}

We have presented DFT plane wave calculations investigating the adsorption preferences of a C$_{60}$ molecule on a Au (111) surface with a protrusion of 4-5 atoms in dependence on an external electric field. We found that the buckyball switches its position from 'on the side' to 'on top' of the Au tip, when the electric field is varied from zero to {\bf E}=$\pm$0.2 V/\AA\ . This is in qualitative agreement with one of the possible intrepretations of recent I/V curve measurements, where a discontinous conductance change at a critical voltage was observed \cite{kubatkin}. In this interpretation the molecule would change its adsorption configuration due to the field and move to the point of highest field strength at the apex of the junction, which in turn would change the distance to the opposite electrode and thereby the conductance. Quantitatively, however, there is a discrepancy of several orders of magnitude beween theory and experiment, where the experimental 'switching voltage' would correspond to only {\bf E}=$\pm$0.0025 V/\AA\ . We argue that this discrepancy could reflect the mismatch in the size of the Au protrusion, where in the experiments it would be much larger and in our simulations it is smaller than the fullerene molecule. In support of this argument we find a substantial decrease in critical voltage, when we increase our tip size by just one Au atom. We also analyzed our results in terms of electrostatics, where field-induced charge transfer between the molecule and the tip was found to be the main physical driving force for the switching effect.

\section*{Acknowledgements}

We are very much indebted to Karsten W. Jacobsen for his encouragement and support throughout this project. We are also grateful for helpful discussions with Jens N{\o}rskov, Jan Rossmeisl, Andrey Danilov and Per Hedeg\aa rd. The Center for Atomic-scale Materials Physics at NanoDTU is sponsored by the Danish National Research Foundation. We acknowledge support from the Nano-Science Center at the University of Copenhagen and from the Danish Center for Scientific Computing through Grant No. HDW-1101-05.

\newpage

\begin{table}
\caption{\label{dipole.tab}Permanent dipole moments $\mu_{perm}$ and polarizabilities $\alpha$ for all geometries in Fig. 1. Induced dipole moments  $\mu_{ind}^{comb}$ and $\mu_{ind}^{isol}$ are also given for an applied field of {\bf E}=-0.25 V/\AA\ , for the combined interface and the sum of dipoles of the isolated fullerene molecule and Au slab, respectively. All values have been calculated directly from planar averages of the charge density. Where appropriate, values calculated from polynomial fits of Figs. 2 and 3 corresponding to Eq. 1 have been added in parantheses.}
\vspace{0.5 cm}
{\centering
\begin{tabular}{c|cccl} \hline
 & \multicolumn{4}{c}{Permanent and induced dipole moments and polarizabilities} \\
geometry & $\mu_{perm}$ [e\AA ]  & $\alpha$ [e\AA$^2$/V] & $\mu_{ind}^{comb}$ [e\AA ] & $\mu_{ind}^{isol}$ [e\AA] \\
\hline
structA/pos1 & -0.617 (-0.662) & 42.742 (41.416) & 11.404 & 5.661 \\
structA/pos2 & -0.520 (-0.546) & 29.766 (29.792) & 7.942  & 5.661 \\
structB/pos1 & -0.845 (-0.923) & 63.852 (56.418) & 14.891 & 6.404 \\
structB/pos2 & -1.168 (-1.289) & 32.302 (32.230) & 7.533  & 6.404 \\
\hline
\end{tabular}\par}\end{table}

\end{document}